\documentclass[traditabstract]{aa} 
\usepackage{graphicx}
\usepackage{amsmath}
\usepackage[USenglish]{isodate}
\usepackage{natbib,twoopt}
\usepackage{hyperref} 
\usepackage{color}
\bibpunct{(}{)}{;}{a}{}{,} 

\newcommand{\Wrad}{$\Omega_{\rm rad}$}
\newcommand{\Wenv}{$\Omega_{\rm env}$}

\defcitealias{LS15}{LS15}

\begin{document} 

%\title{Constraints on angular momentum transport in main sequence solar-like stars from observed rotation periods in open clusters}
\title{On the competing effect of wind braking and interior coupling in the rotational evolution of solar-like stars}

\titlerunning{Rotational coupling in solar-like stars}

\author{
F. Spada\inst{1}
\and
A.~C. Lanzafame\inst{2,3}
}

\authorrunning{F. Spada \& A.~C. Lanzafame}

\offprints{F. Spada \\ \email{spada@mps.mpg.de}}

\institute{
Max-Planck-Institut f\"ur Sonnensystemforschung, Justus-von-Liebig-Weg 3, 37077 G\"ottigen, Germany 
\and 
Universit\`a di Catania, Dipartimento di Fisica e Astronomia, Sezione Astrofisica, Via S. Sofia 78, 95123 Catania, Italy 
\and
INAF-Osservatorio Astrofisico di Catania, Via S. Sofia 78, 95123 Catania, Italy
}

\date{}

\abstract{
Solar-like stars ($M \lesssim 1.3\, M_\odot$) lose angular momentum through their magnetized winds.
The resulting evolution of the surface rotation period, which can be directly measured photometrically, has the potential to provide an accurate indicator of stellar age, and is constrained by observations of rotation periods of coeval stars, such as members of Galactic open clusters.
A prominent observational feature of the mass--rotation period diagrams of open clusters is a sequence of relatively slower rotators.
The formation and persistence of this slow rotators sequence across several billion years imply an approximately coherent spin-down of the stars that belong to it.
In particular, the sequence is observed to evolve coherently towards longer periods in progressively older clusters. \\
Recent observations of the $\approx 700$-Myr old Praesepe and the $1$-Gyr old NGC 6811 clusters, however, seem to contradict this general pattern.
While the $1\, M_\odot$ stars on the slow rotators sequence of the older NGC 6811 have longer periods than their counterparts in the younger Praesepe, as expected, the two sequences essentially merge at lower masses ($\lesssim 0.8\, M_\odot$).
In other words, low-mass stars seem to have not been spinning down at all in the intervening $300$ Myr. \\
Here we show that this behavior is a manifestation of the variable rotational coupling in solar-like stars.
The resurfacing of angular momentum from the interior can temporarily compensate for the loss due to wind braking at the surface.
In our model, the internal redistribution of angular momentum has a steep mass dependence; as a result, the re-coupling occurs at different ages for stars of different masses.
The semi-empirical mass-dependence of the rotational coupling timescale included in our model produces an evolution of the slow rotators sequence in very good agreement with the observations.
Our model, in particular, explains the stalled surface spin-down of low-mass stars between Praesepe and NGC 6811, and predicts that the same behavior should be observable at other ages in other mass ranges.
}


\maketitle

\section{Introduction}
\label{intro}

% OPENING
% general introduction
Solar-like stars  ($M \lesssim 1.3\, M_\odot$) have outer convection zones where magnetic fields are generated through dynamo action. 
These stars have therefore magnetically active atmospheres, and the braking torque exerted by their magnetized winds efficiently carries away angular momentum from their surfaces \citep[e.g.,][]{Schatzman:1962, Kraft:1967, Weber_Davis:1967, Kawaler:1988}.  
As a result, in contrast to their more massive counterparts, solar-like stars undergo a significant rotational evolution on the main sequence.
For instance, the rotation period of a star of solar mass is observed to be $P_{\rm rot} \approx 1$ day upon reaching the zero-age main sequence ($40$ Myr), to be contrasted with the $26.09$ days of the present Sun ($4.57$ Gyr).
The rotation period of solar-like stars, moreover, can be derived accurately from the analysis of their photometric light curves.
The availability of a directly measurable quantity undergoing such large variations on the main sequence holds great potential as an accurate age indicator \citep[``gyrochronology":][]{Barnes:2003, Barnes_ea:2016b}.

%%% THE SRS
% observations in clusters and SRS
The most stringent constraints on the rotational evolution of solar-like stars come from $P_{\rm rot}$ measurements for stars in Galactic open clusters, whose age is independently known from classical methods, such as isochrone fitting \citep[e.g.,][]{Demarque_Larson:1964}.
The mass--rotation period diagram of a cluster, in analogy to his classical counterpart, the color-magnitude diagram, reveals the (rotational) evolutionary state of a sample of stars of different mass at a fixed age.

While stars in early pre-main sequence clusters (age $\approx 1$--$10$ Myr, see, e.g., \citealt{Moraux_ea:2013}) have a broad range of periods with an approximately uniform distribution in mass, in older clusters a clear pattern gradually emerges.
A distinct sequence of relatively slower rotators is clearly visible in $\approx 70$ Myr-old clusters, or older \citep[see Figure 1 of][]{Barnes:2003}.

At its outset, the slow rotators sequence is well-defined only at the high mass end of the solar-like regime ($M \approx 1.3 \, M_\odot$), while at lower masses it coexists with a less structured broad distribution of faster rotators.
The fast rotators gradually disappear in clusters of increasing age, as stars of lower and lower mass converge to the slow rotators sequence. 
By the age of $1$ Gyr, all stars down to $0.6 \, M_\odot$ have joined the slow rotators sequence.
Although observations suggest that stars of mass $M < 0.6 \, M_\odot$ eventually reach the slow rotators sequence as well, data in sufficiently old cluster and for low masses are still too scarce to draw definitive conclusions \citep[but see][and references therein]{Newton_ea:2018}.

% open problem of how the sequence forms; M dwarfs: hic sunt leones
The details of the formation of the slow rotators sequence are still not well understood, and only semi-empirical models have been proposed so far \citep{Barnes:2010, Brown:2014, Gondoin:2017}.
The observational evidences summarized above seem to imply that stars settle down on the slow rotators sequence by means of a quick, one-time transition, with a mass-dependent timescale.
Since stars of the same mass do not all converge on the slow rotators sequence at once, a third parameter (besides mass and age), and/or a stochastic transition from the fast- to the slow-rotator regime must be involved.

% mention skumanich-like evolution of the SRS
Beside the open issue of its mechanism of formation, the persistence of the slow rotators sequence (clearly observed up to at least $\approx 4$ Gyr in M 67: \citealt{Barnes_ea:2016a}) points to an intrinsic regularity in the rotational evolution of stars once they have converged on it.
The first attempts to model the evolution of the  slow rotators sequence assumed its mass dependence to be factorizable from the age dependence, and that the latter follows the classical spin-down relation proposed by \citet{Skumanich:1972}, $P_{\rm rot}\propto \sqrt{t}$ \citep{Barnes:2003, Barnes:2007}.  
Although these assumptions were later found to be inadequate (\citealt{Meibom_ea:2009, Meibom_ea:2011, Barnes_Kim:2010,LS15}, hereafter \citetalias{LS15}), the slow rotators sequences of clusters of different ages follow a monotonic relation in the mass-rotation period diagram, with older clusters corresponding to slower periods. 

\begin{figure}
\begin{center}
\includegraphics[width=0.5\textwidth]{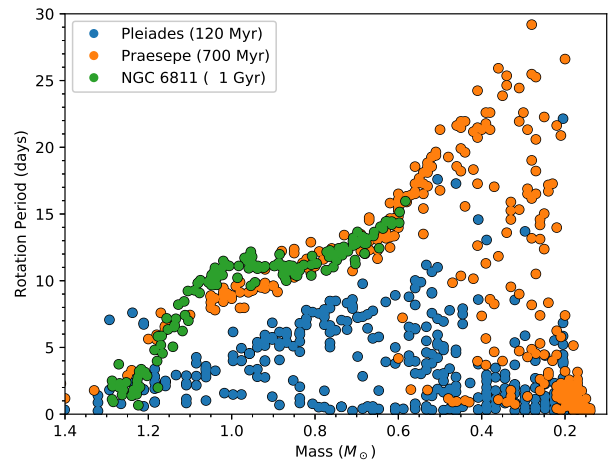}
\caption{Emergence and evolution of the slow rotators sequence in the mass-period diagram of open clusters, from the $120$ Myr-old Pleiades to the $1$ Gyr-old NGC $6811$. Data sources: Pleiades: \citet{Rebull_ea:2016}; Praesepe: \citet{Douglas_ea:2019}; NGC 6811: \citet{Curtis_ea:2019}.}
\label{fig:data}
\end{center}
\end{figure}

%%% the problem: new data on rotation in the Pleiades, Praesepe, and NGC 6811 seem to imply that K stars undergo an epoch of ``stalled rotational braking".
The most recent observations of the Praesepe cluster ($700$ Myr) and NGC 6811 clusters ($1$ Gyr), featuring an unprecedented sampling of the low-mass regime of their slow rotators sequences, seem to contradict this (so far) well-established fact \citep{Curtis_ea:2019,Douglas_ea:2019}.
While $\approx 1\, M_\odot$ stars on the slow rotators sequence of these two clusters display the expected behavior (i.e. the younger Praesepe stars rotating faster than those in the older NGC 6811), the two sequences seem to overlap below $0.8\, M_\odot$ (see Figure \ref{fig:data}).
In other words, low-mass stars in NGC 6811 seem to have not been spinning down significantly in the intervening $300$ Myr. 
%and appear younger in a gyrochronological sense.

%%% OUR MODEL
In this paper, we model the evolution of the slow rotators sequence, and in particular we provide a theoretical interpretation of the puzzling new data from Praesepe and NGC 6811.
We show that our model, based on our previous work \citepalias{LS15}, can reproduce the apparent stalled spin-down observed in NGC 6811, and that this is the result of the redistribution of angular momentum from the stellar interior to the surface, which is temporarily able to offset the effect of the magnetic wind braking.
The two key ingredients of the model are the wind-braking law and the mass-dependence of the rotational coupling timescale originally proposed by \citetalias{LS15}.
The latter is also in remarkable agreement with the results of \citet{Somers_Pinsonneault:2016}, obtained from an independent analysis of a related, but distinct problem (the lithium depletion in solar-like stars). 
%Our model is a promising step towards a physically motivated description of the slow rotators sequence, beyond the semi-empirical formulations proposed so far.

%%% IMPLICATIONS
% mention open problem of angular momentum transport in the interior (REFS)
Our results provide valuable constraints on the mechanisms that transport angular momentum in the interior of solar-like stars, whose physical nature is still uncertain and strongly debated \citep[e.g.,][]{Charbonneau_MacGregor:1993, Ruediger_Kitchatinov:1996, Talon_Charbonnel:2005, Spada_ea:2010, Eggenberger_ea:2019}.
Moreover, we demonstrate that the assumption of a solid-body rotation profile in stellar interiors is inadequate to obtain accurate gyrochronology relations.

This paper is organized as follows: we describe the observational data used to constrain our models in Section \ref{data}; we outline the physics of our rotational evolution model in Section \ref{model}; our results are presented in Section \ref{results}; we discuss our findings in Section \ref{discussion}; our conclusions are summarized in Section \ref{conclusions}.

\section{Observed evolution of the slow rotators sequence from 120 Myr to 1 Gyr}
\label{data}

We constrain our models with the most up-to-date observations available for three rich, well-studied clusters, namely, the Pleiades, Praesepe, and NGC 6811 \citep[][respectively]{Rebull_ea:2016,Douglas_ea:2019,Curtis_ea:2019}.
In the mass-rotation period diagram of each of these clusters the slow rotators sequence is clearly recognizable between $M \approx 0.3$ and $1.3\, M_\odot$, encompassing the entire domain of solar-like stars (see Figure \ref{fig:data}). 

The Pleiades is one of the youngest clusters in which the slow rotators sequence is observed; a large fraction of stars have not yet converged onto it.
After approximately half a billion year, in the Praesepe cluster, most of the stars of mass $\gtrsim 0.6\, M_\odot$ are on the slow rotators sequence.
Finally, in NGC 6811, at the age of $1$ Gyr, all stars with measured rotation period are on the sequence.

As can be seen in Figure \ref{fig:data}, the slow rotators sequences of Praesepe and NGC 6811 overlap below $M \lesssim 0.8\, M_\odot$. 
This behavior is at odds with a simple description of the slow rotators sequence as coherently evolving in the mass-rotation period diagram.
Indeed, stars of mass $\approx 0.7\, M_\odot$ seem to undergo a ``stalled spin-down'' epoch between $700$ Myr and $1$ Gyr, while their more massive counterparts spin down as expected \citep{Curtis_ea:2019}.

\section{A two-zone rotational evolution model with mass-depedent coupling}
\label{model}

\subsection{Description of the model}

We apply the rotational evolution model formulated by \citetalias{LS15} to describe the evolution of the slow rotators sequence.
The main assumption of the two-zone model is that the radiative zone and the convective envelope of the star are in a state of rigid rotation, with angular velocities $\Omega_{\rm rad}$ and $\Omega_{\rm env}$, respectively \citep{MacGregor_Brenner:1991}.
These two quantities specify completely the rotational state of the star at a given time; the two zones have therefore angular momenta $J_{\rm rad} = I_{\rm rad} \, \Omega_{\rm rad}$, and $J_{\rm env} = I_{\rm env}\, \Omega_{env}$, respectively, $I_{\rm rad}$ and $I_{\rm env}$ being their moments of inertia.

The physics included in the model is as follows:
\begin{enumerate}
%%%
\item{\it Initial conditions.}
We assume initial rigid rotation, $\Omega_{\rm rad}(t_0)=\Omega_{\rm env}(t_0)$, with $t_0 \approx 1$ Myr. 
The initial period is assumed to be $P_0 = 8$ days for all masses; this choice is in qualitative agreement with the observed rotation period distribution of very young clusters (e.g., the $1$--$4$ Myr-old Orion Nebula Cluster, \citealt{Rebull:2001}; see also \citealt{Moraux_ea:2013}). 
The interaction with the circumstellar disk is taken into account according to the disk-locking hypothesis \citep{Koenigl:1991}, i.e., assuming that \Wenv{} remains constant for the duration of the disk lifetime.
We assume a circumstellar disk lifetime of $5$ Myr, independent of mass \citep{Hernandez_ea:2008}.
%%%
\item{\it Wind braking.}
The overall rotational evolution is driven by the angular momentum loss from the surface, due to the torque imposed by the magnetized stellar wind \citep[e.g.,][]{Schatzman:1962,Weber_Davis:1967,Kawaler:1988}. 
We adopt the wind braking law originally formulated by \citetalias{LS15}, which incorporates the mass dependence proposed by \citet{Barnes_Kim:2010}, and follows the classical $\propto \Omega_{\rm env}^3$ rotation rate dependence of \citet{Kawaler:1988}: 
\begin{equation}
\label{wind_braking}
\dot J_{\rm wb} =  
- K_w \, K_0 \left(\frac{I_{\rm star} \tau_{\rm ov}}{{I_{\rm star, \odot} \tau_{\rm ov, \odot}}} \right) \Omega_{\rm env}^3 .
\end{equation}
In the equation above, $I_{\rm star}=I_{\rm rad}+I_{\rm env}$ is the moment of inertia of the whole star, and $\tau_{\rm ov}$ is the convective overturn timescale of the convection zone. 
The product $I_{\rm star} \, \tau_{\rm ov}$ encompasses the mass dependence of our wind braking law; both quantities in the product are normalized to their values for a $1\, M_\odot$ model, as indicated by the subscript ``$\odot$''; $K_w$ is an overall calibration constant, whose scaling $K_0 = 1.11 \times 10^{47}$ g cm$^2$ s depends on the choice of the units. 
%%%
\item{\it Internal angular momentum transport.}
As the envelope loses angular momentum, differential rotation develops, i.e., $\Omega_{\rm rad}(t) \gtrsim \Omega_{\rm env}(t)$; exchange of angular momentum between the two zones can also occur. 
During the pre-main sequence phase, the radiative interior grows at the expense of the convection zone, which comprises the whole star during the Hayashi phase \citep[e.g.,][]{KWW12}. 
Once on the main sequence, angular momentum redistribution between the radiative and the convective zone can be mediated by several processes, whose relative importance is still an open issue \citep[see][for a recent review]{Bouvier_ea:2014}. 
Our model accounts for this effect phenomenologically, by introducing a constant, mass-dependent timescale $\tau_c$, over which the excess of angular momentum of the interior, $\Delta J \equiv \dfrac{I_{\rm env} J_{\rm rad} - I_{\rm rad} J_{\rm env}}{I_{\rm rad}+I_{\rm env}}$, is transferred to the envelope.
The rotational coupling timescale $\tau_c$ is taken to be constant along the evolution, and to scale with the stellar mass as:
\begin{equation}
\label{tauc}
\tau_c = \tau_{c, \odot} \cdot \left(\frac{M_*}{M_\odot}\right)^{-\alpha}.
\end{equation}
This scaling, derived from semi-empirical fitting of \citetalias{LS15}, was found by these authors to be robust to the choice of the wind braking law.
It was also found to be in remarkably good agreement with the independent analysis of \citet{Somers_Pinsonneault:2016}.
\end{enumerate}

The two-zone model equations for the the evolution of \Wrad{} and \Wenv{} are: 
\begin{align}
\nonumber
& I_{\rm rad} \dot \Omega_{\rm rad} = + \dfrac{2}{3} \dot M_{\rm rad} R_{\rm rad}^2 \,\Omega_{\rm env} - \dfrac{\Delta J}{\tau_c}  - \dot I_{\rm rad} \Omega_{\rm rad};
\\
\label{tzm}
\\
\nonumber
& I_{\rm env} \dot \Omega_{\rm env} = - \dfrac{2}{3} \dot M_{\rm rad} R_{\rm rad}^2 \, \Omega_{\rm env} + \dfrac{\Delta J}{\tau_c}   - \dot I_{\rm env} \Omega_{\rm env} + \dot J_{\rm wb},
\end{align}
where the dot denotes differentiation with respect to time; $R_{\rm rad}$  and $M_{\rm rad}$ are the radius and the mass of the radiative core, respectively. 

All the stellar structure parameters ($M_{\rm rad}$, $R_{\rm rad}$, $I_{\rm rad}$, $I_{\rm env}$, $\tau_{\rm ov}$) are derived from stellar evolution models constructed with the YREC code in its non-rotational configuration \citep{Demarque_ea:2008}.

Our model contains five parameters in total: the initial conditions are set by the initial rotation period $P_0$ and by the circumstellar disk lifetime $\tau_{\rm disk}$; the wind braking law contains the calibration constant $K_w$; the rotational coupling timescale is specified by $\tau_{c, \odot}$ and $\alpha$.

\subsection{Comparison with \citetalias{LS15} and re-determination of the model parameters}

\citetalias{LS15} presented a comparison of several wind braking laws and a statistical determination of the parameters of the two-zone model ($P_0$, $\tau_{\rm disk}$, $\tau_c$), based on the observational constraints on the slow rotators sequence available at the time.
The best-fitting values of the model parameters were obtained by means of a Markov Chain Monte Carlo (MCMC) procedure for several different stellar masses in the range $M \approx 0.7$--$1.1\, M_\odot$ (see \citetalias{LS15} for details). 
The main conclusions of their analysis can be summarized as follows.
First, the wind braking law \eqref{wind_braking} (referred to as ``KB" in \citetalias{LS15}) captures the observed shape of the slow rotators sequence sufficiently well that it can be used to describe the rotational evolution of stars of different masses with a unique value of the calibration constant $K_w$.
Second, moving from solar- towards low-mass stars, it is increasingly important to take into account internal differential rotation in order to accurately reproduce the observations; the best results were obtained with a rotational coupling timescale that scales with stellar mass according to equation \eqref{tauc}. 

Building on our previous work, in this paper we adopt the two-zone model \eqref{tzm}, and we retain the functional forms of the wind braking law \eqref{wind_braking} and of the mass-dependent coupling timescale \eqref{tauc}.
However, we re-determine the parameters to take into account the new data at lower masses available for Praesepe and NGC 6811.
In particular, our goal is to ascertain whether our model can reproduce the stalled spin-down of low-mass stars observed in the slow rotators sequence of NGC 6811.

%The values of the parameters adopted in this work are listed and compared to those of \citetalias{LS15} in Table \ref{tab:parameters}.
%\begin{table}
%\caption{Best-fitting parameters of the two-zone model.}
%\begin{center}
%\begin{tabular}{lcc}
%\hline
%\hline
%Parameter & This Work & \citetalias{LS15} (KB) \\
%\hline
%$P_0$    (days)               &    $8.0$    &  $8.0$    \\  
%$\tau_{\rm disk}$ (Myr)   &    $5.0$    &  $5.0$    \\ 
%$K_w$                             &   $3.2$    &   $4.5$    \\
%$\tau_{c, \odot}$  (Myr)   &    $25.$   &   $25.$    \\
%$\alpha$                          &   $5.2$    &   $7.3$    \\
%\hline
%\end{tabular}
%\tablefoot{}
%\end{center}
%\label{tab:parameters}
%\end{table}

\begin{figure}
\begin{center}
\includegraphics[width=0.5\textwidth]{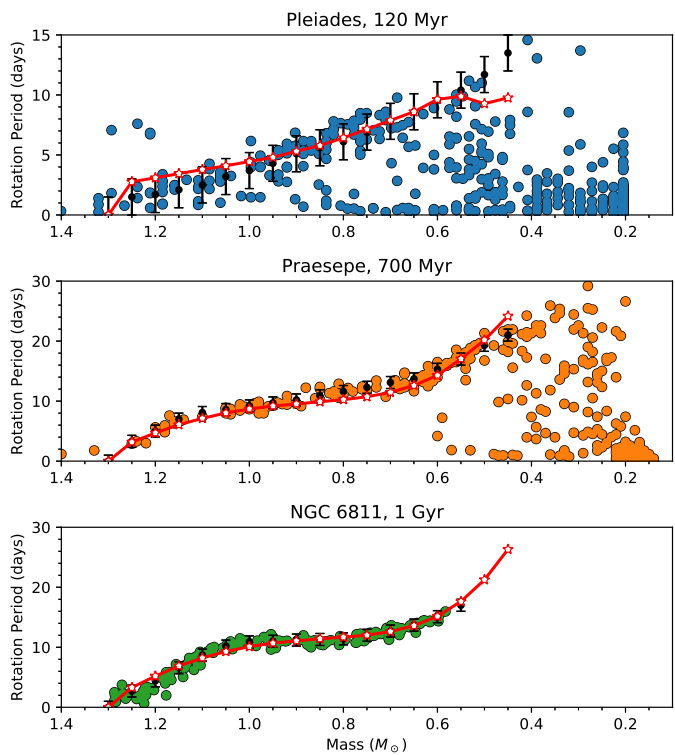}
\caption{Synthetic slow rotators sequences predicted by our two-zone model at $120$ Myr, $700$ Myr, and $1$ Gyr (red solid lines with stars) compared with the observations (colored circles). The black circles with error bars mark the non-parametric fits of the slow rotators sequences; the error bars shown are $1.5$ days for the Pleiades' sequence and $1$ day for Praesepe and NGC 6811.}
%Model parameters: $P_0=8$~days, $\tau_{\rm disk}=5$~Myr, $K_w=3.4$, $\tau_{c, \odot}=22$~Myr, $\alpha = 5.6$. Data as in Figure \ref{fig:data}.}
\label{fig:sequence}
\end{center}
\end{figure}

\begin{figure}
\begin{center}
\includegraphics[width=0.5\textwidth]{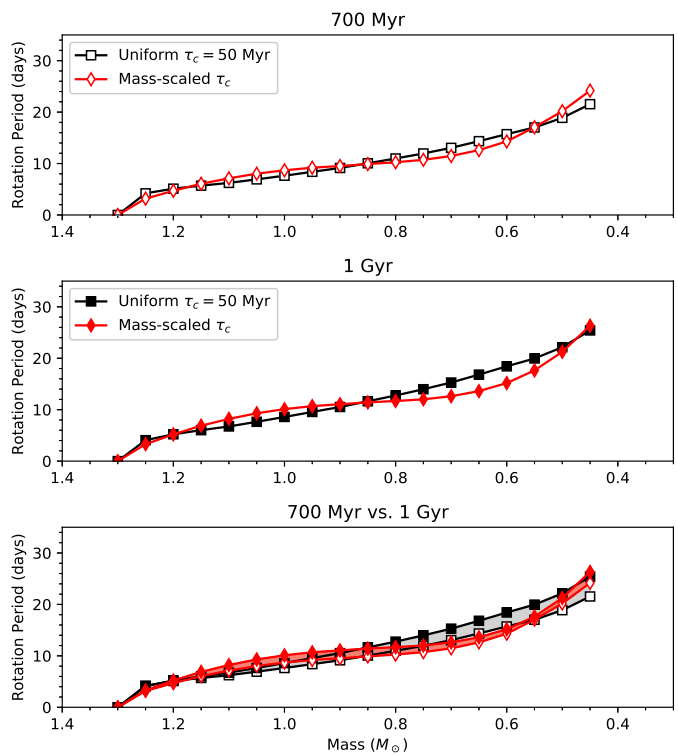}
\caption{Synthetic slow rotators sequences obtained with uniform $\tau_c$ vs. mass-dependent $\tau_c$. 
The bottom panel shows a comparison of the difference between the slow rotators sequences at $700$ Myr and $1$ Gyr (gray and red shading for the uniform $\tau_c$ and the mass-dependent $\tau_c$, respectively).}
\label{fig:compare}
\end{center}
\end{figure}

The parameters governing the initial conditions, $P_0$ and $\tau_{\rm disk}$, are determined by the requirement to reproduce the slow rotators sequence of the Pleiades, and have the same values as in \citetalias{LS15}. 

The parameters $K_w$, $\tau_{c, \odot}$, and $\alpha$, on the other hand, control the subsequent evolution of the slow rotators sequence from $\approx 100$ Myr onward.
These parameters are strongly correlated with each other \citepalias[see][]{LS15}, and must therefore be re-evaluated together.  
The new data on Praesepe and NGC 6811, extending significantly the mass range sampled, provide the opportunity to refine the estimate of these parameters, in particular the power law exponent $\alpha$. 

The revised best-fitting values of the parameters $K_w$, $\tau_{c, \odot}$, and $\alpha$ were obtained using a Python implementation of the \citet{Nelder_Mead:1965} simplex algorithm, which is available in the function \texttt{optimize.minimize}, part of the \texttt{SciPy}\footnote{http://www.scipy.org/} package.
Our revised estimates are:
\begin{equation}
K_w = 3.4; \ \ \ \  \tau_{c,\odot} = 22\, {\rm Myr}; \ \ \ \ \alpha = 5.6. 
\end{equation}

It should be emphasized that these parameters were not obtained by \citetalias{LS15} as a direct result of their MCMC procedure. 
Rather, $K_{\rm w}^{\rm LS15}=4.5$ was obtained as an average over the entire mass range considered ($\approx 0.7$--$1.1\, M_\odot$), while $\tau_{c, \odot}$, and $\alpha$ were determined a posteriori from the values of $\tau_c$ fitted independently at each mass.
Our current estimate of $K_w$ is compatible with the \citetalias{LS15} value within the uncertainties, and is in excellent agreement with their estimate for a $1\, M_\odot$ star ($3.6$, cf. Table 3 of \citetalias{LS15}). 
The value of $\tau_{c, \odot}$ is nearly unchanged ($\tau_{c, \odot}^{\rm LS15}=25$ Myr).
Finally, the re-calibrated $\alpha$ is significantly different from $\alpha^{\rm LS15} = 7.3$.
This difference is a consequence of the extension to a wider mass range, as well as the intrinsic ill-conditioning of the best-fitting procedure, due to the strong correlations among the parameters of the model.

\section{Results}
\label{results}

\subsection{Modeling the evolution of the slow rotator sequence}

The results of our best-fitting two-zone model are compared with the observations in Figure \ref{fig:sequence}.
A non-parametric fit of the slow rotators sequences, constructed consistently with the procedure employed by \citetalias{LS15}, is also shown for each cluster (black circles with error bars). 
Error bars of $1.5$ days for the Pleaides, and of $1$ day for Praesepe and NGC 6811, represent the intrinsic scatter of the sequences, which is astrophysical in origin.

The overall fit of the slow rotators sequences of the Pleiades, Praesepe, and NGC 6811 is quite good (Figure \ref{fig:sequence}).
In particular, our model reproduces the apparently stalled surface spin-down of the $\approx 0.7\, M_\odot$ stars between $700$ Myr and $1$ Gyr with respect to their more massive counterparts, which results in the merging of the slow rotators sequences of Praesepe and NGC 6811.

The key ingredient in reproducing this behavior is the scaling of the rotational coupling timescale with the mass of the star according to equation \eqref{tauc}.
This is immediately apparent when we compare the predictions of our reference model with those obtained assuming a value of $\tau_c$ independent of stellar mass (all the other parameters being the same).

As Figure \ref{fig:compare} shows, the uniform-$\tau_c$ assumption produces a slow rotators sequence with a steeper slope at the age of NGC 6811 in the mass range $\approx 0.5$--$0.85\, M_\odot$.  
Moreover, the difference between the slow rotators sequences at $700$ Myr and at $1$ Gyr (bottom panel of the Figure) is monotonically increasing towards lower masses for the  uniform-$\tau_c$ model.
We conclude that a model with $\tau_c$ independent of mass cannot achieve a satisfactory fit of the slow rotators sequences of Praesepe and NGC 6811. 
Such a model, in particular, fails to reproduce the stalled spin-down of the $\approx 0.7\, M_\odot$ stars.

On the contrary, the two-zone model implementing the scaling of $\tau_c$ according to equation \eqref{tauc} produces a separation between the slow rotators sequences at $700$ Myr and $1$ Gyr that is largest at $\approx 1 \, M_\odot$ and decreases at lower masses, thus correctly reproducing the merging of the slow rotators sequences observed in Praesepe and NGC 6811.

\begin{figure}
\begin{center}
\includegraphics[width=0.5\textwidth]{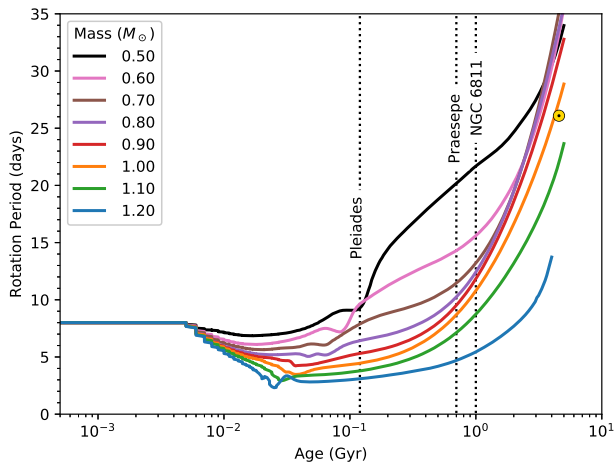}
\caption{Rotational evolution tracks constructed with our two-zone model. The vertical dotted lines indicate the ages of the clusters used for comparison (cf. Figure \ref{fig:sequence}); the Sun is also plotted for reference.}
\label{fig:tracks}
\end{center}
\end{figure}

Figure \ref{fig:tracks} shows the evolution of the surface rotation period calculated with our rotational evolution model. 
This alternative view further illustrates the effect of the rotational coupling.
Stars of mass $\lesssim 0.8 \, M_\odot$ experience a significantly reduced surface spin-down than their more massive counterparts between $700$ Myr and $1$ Gyr.
This effect is the direct consequence of the redistribution of angular momentum from the radiative zone to the convective envelope, which temporarily offsets the angular momentum loss at the surface.

\begin{figure}
\begin{center}
\includegraphics[width=0.5\textwidth]{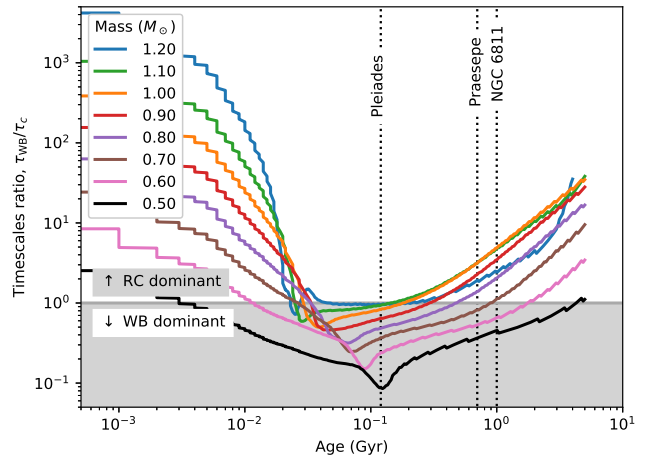}
\caption{Ratio of the wind braking timescale $\tau_{\rm wb}$ over the rotational coupling timescale $\tau_c$ for the same tracks shown in Figure \ref{fig:tracks}. The shaded area of the plot marks the evolutionary phases where wind braking (WB) is dominant over the rotational coupling (RC; see the discussion in the text).}
\label{fig:timescales}
\end{center}
\end{figure}

\subsection{Wind braking vs. rotational coupling}

The rotational evolution according to the two-zone model equations \eqref{tzm} is controlled by the wind braking, described by equation \eqref{wind_braking}, and by the rotational coupling, characterized by equation \eqref{tauc}.
Which of these two processes dominates over the other depends on the mass of the star as well as its age.

We can gain some insight into the relative importance of these two effects by comparing their timescales.
The wind braking law \eqref{wind_braking} has a steep dependence on the surface rotation rate ($\propto \Omega_{\rm env}^3$). 
The efficiency of the angular momentum loss therefore varies by several orders of magnitude between the zero-age main sequence, when the rotation rate is maximum for all masses, and the mature main sequence phases ($\gtrsim 1$ Gyr).
We can characterize the wind braking efficiency by the timescale:
\begin{equation}
\frac{1}{\tau_{\rm wb}} \equiv \frac{\dot J_{\rm wb}}{J_{\rm env}}.
\end{equation}
The rotational coupling timescale, on the other hand, is assumed to be constant for the entire evolution of the star.

The ratio $\tau_{\rm wb}/\tau_c$ is a measure of the relative importance of the angular momentum loss and the internal angular momentum transport. 
This ratio is plotted in Figure \ref{fig:timescales}.
When $\tau_{\rm wb}/\tau_c < 1$, the angular momentum redistribution from the interior to the surface is not efficient enough to compensate for the angular momentum lost from the surface, and significant differential rotation develops. 
The region of the plot in Figure \ref{fig:timescales} where this condition is realized is shaded in gray. 
On the contrary, when $\tau_{\rm wb}/\tau_c > 1$, the rotational coupling is able to maintain a quasi solid-body rotation profile inside the star, while its angular momentum evolution is driven by the wind braking at the surface.

For stars of solar mass or higher, the rotational coupling catches up with the wind braking already by the age of the Pleiades, and differential rotation has essentially no chance to develop at later times.
Conversely, for stars of mass $\approx 0.75 \, M_\odot$ the angular momentum transport from the interior is still less efficient than the wind braking up to $\approx 700$ Myr. 
For a $0.5\, M_\odot$ star, the rotational coupling is still relatively inefficient even after several Gyr.
The evolution of the surface rotation period of stars less massive than $\approx 0.75 \, M_\odot$ will therefore be affected by the angular momentum resurfacing from the radiative interior between the age of Praesepe and that of NGC 6811.

From Figure \ref{fig:timescales} we can also deduce that the same effect should be observed in other mass ranges at other ages.
For instance, our model predicts that this angular ``momentum reservoir" effect is operational for stars of mass $\approx 0.8$--$1.0 \, M_\odot$ at an age of $\approx 300$ Myr, and essentially up to the age of the Sun for $M \lesssim 0.6 \, M_\odot$.
We look forward to testing this prediction as rotation periods for cluster stars of appropriate age and mass become available in the future.

\section{Discussion}
\label{discussion}

We have presented a simple model for the rotational evolution of solar-like stars, which contains two main physical ingredients: the magnetized wind braking at the surface, and the angular momentum transport (or ``coupling'') between the interior and the surface.

The essential role of the rotational de-coupling and re-coupling between the radiative interior and the convective envelope in the pre-main sequence and early main sequence evolution of solar-like stars has been recognized since the works of \citet{Endal_Sofia:1981, Stauffer_ea:1984, Stauffer_ea:1985, Pinsonneault_ea:1989, Soderblom_ea:1993}, just to name a few.
This effect has been incorporated phenomenologically in all subsequent modeling efforts, and several candidate processes have been proposed (MHD waves/instabilities, gravity waves, see, e.g., \citealt{Charbonneau_MacGregor:1993, Ruediger_Kitchatinov:1996, Spruit:2002, Talon_Charbonnel:2003, Talon_Charbonnel:2005, Spada_ea:2010, Brun_ea:2011, Oglethorpe_Garaud:2013}).
In spite of this, the physical nature of the processes redistributing angular momentum in the interior of stars is still a major open question. 

Previous works \citep[e.g.,][]{Denissenkov_ea:2010, Gallet_Bouvier:2015} have underlined the strong mass dependence of the core-envelope rotational coupling timescale. 
By concentrating on the rotational evolution of the slow rotators sequence (or ``I-sequence'' in the terminology of \citealt{Barnes:2003}), \citetalias{LS15} quantified the mass dependence of the rotational coupling timescale. 
The model proposed by \citetalias{LS15}, which incorporates such dependence, can reproduce accurately the evolution of the slow rotators sequence.
The new rotational data in NGC 6811 \citep{Curtis_ea:2019} provide previously unavailable constraints, by extending the range of mass with known rotation periods at $1$ Gyr down to $M \approx 0.6 \, M_\odot$. 
Together with the new Praesepe data \citep{Douglas_ea:2019}, these prompted us to re-evaluate the parameters of the \citetalias{LS15} model.

The re-calibrated parameters are consistent with the results of \citetalias{LS15} within the uncertainties of their fit to the data available at that time. 
Our updated model also explains quantitatively the apparent halt in the spin-down of low-mass stars ($M\approx 0.7\, M_\odot$) between $700$ Myr and $1$ Gyr, recently discovered by \citet{Curtis_ea:2019}. 
Indeed, this phenomenon is a manifestation of the angular momentum redistribution from the interior of the star to its outer envelope, which can temporarily compensate for the angular momentum lost from the surface via magnetic braking. 

It should be emphasized that the observed lack of spin-down arises naturally in our model from simple assumptions regarding the mass-dependence of the angular momentum coupling timescale.
In other words, it is not necessary to postulate a phase of weakened wind braking regime to explain the evolution of the slow rotators sequence as observed in open clusters.
In addition, our model provides quantitative predictions on the duration of this suspended spin-down phase, and on the epoch at which stars of different mass are expected to experience it.
These predictions will be testable as new rotational data, sampling the slow rotators sequence at low stellar masses and in clusters of different ages, become available.

\citet{Barnes:2003} originally defined the I-sequence observed in open clusters as comprising stars spinning down according to the phenomenological relation $P_{\rm rot} \propto \sqrt{t}$ \citep{Skumanich:1972}.
Our results show that such a working definition still retains its approximate validity even in the light of the most recent observations.
Indeed, the overall shape of the slow rotators sequence is captured remarkably well by the wind braking law \eqref{wind_braking}, which couples the mass-dependence proposed by \citet{Barnes_Kim:2010} with a dependence on $\Omega_{\rm env}^3$, which is well-known to reproduce the \citeauthor{Skumanich:1972} law \citep[see, e.g.,][]{Kawaler:1988}.

Departures from the \citeauthor{Skumanich:1972} law in the slow rotators sequence have been know for a long time \citep[e.g.,][]{Meibom_ea:2009, Meibom_ea:2011}, and were already interpreted by \citetalias{LS15} as arising from the re-coupling of the surface with the interior.
This effect manifests a strong dependence of the coupling timescale on stellar mass (cf. equation \ref{tauc}).
Building on our previous results, we show that the halt in the spin-down of low-mass stars reported by \citet{Curtis_ea:2019} is a consequence of the re-coupling occurring at the age of $\lesssim 1$ Gyr for $\approx 0.7\, M_\odot$ stars.

The new observations of \citet{Curtis_ea:2019} allow us to refine the determination of the parameters in equation \eqref{tauc}, to extend its validity to lower masses.
This places valuable constraints on the unknown physical nature of the processes that transport angular momentum in the interior of solar-like main sequence stars.
In particular, we confirm (within the uncertainties) the steep mass-dependence found by \citetalias{LS15}, and independently recovered by \citet{Somers_Pinsonneault:2016}. 
A successful theoretical description of the rotational coupling in solar-like stars from first principles should be able to explain the semi-empirical result of equation \eqref{tauc}.

The deviations from a purely \citeauthor{Skumanich:1972} spin-down that depend on the stellar mass are of practical importance for gyrochronology.
Indeed, the original gyrochronological relations, based on the assumption of a purely factorable color and age dependence (where the color is used as a proxy of stellar mass), $P_{\rm rot} = f(B-V)\, g(t)$, were already shown to be untenable by \citet{Barnes_Kim:2010}.
Our model is a promising step towards the construction of more accurate gyrochronology relations, extracted from a simple physical model.
This improvement comes at the price of not having a simple all-encompassing formula that can be readily inverted to derive the age from the rotation period.
For convenience of use, we provide updated rotational isochrones for a wide range of ages in Appendix \ref{appendix}.

%%% LIMITATIONS of the model
%Our model is not yet applicable to describe the approach and convergence to the slow rotators sequence. 
%This process is likely to transpire as a rapid transient phase, which has been speculatively linked to a re-organization of the magnetic configuration of the star, but for which only phenomenological descriptions are currently available \citep[e.g.,][]{Barnes:2010, Brown:2014, Gondoin:2017}.

Our model reproduces satisfactorily the slow rotators sequence as observed in the Pleiades cluster ($\approx 100$ Myr), and its subsequent evolution, as constrained by the recent observations of Praesepe and NGC 6811 ($700$ Myr and $1$ Gyr, respectively), in the entire domain of solar-like stars ($\approx 0.4$--$1.3 \, M_\odot$).
A two-zone model like ours, however, is obviously not applicable without modifications to fully convective stars ($\lesssim 0.35 \, M_\odot$).
Whether such low-mass stars converge on the slow rotators sequence, and whether their evolution is similar to that of their more massive counterparts in spite of this qualitative difference in their interior structure, are still open questions.

\section{Conclusions}
\label{conclusions}

We have presented a model for the rotational evolution of solar-like stars that reproduces the features of the slow rotators sequence as observed in open clusters between $\approx 100$ and $1000$ Myr. 
In particular, our model captures satisfactorily its mass-dependence in the range $0.4$--$1.3\, M_\odot$, and the apparent stalled spin-down observed in $\approx 0.7\, M_\odot$ stars between $700$ Myr and $1$ Gyr.

The key ingredients of the model are the scalings of the wind-braking law and of the rotational coupling timescale with stellar mass.
The former is well-represented by the product of the moment of inertia of the star times its convective overturn timescale; the latter follows a steep power-law (of exponent $\approx -5.6$), with a coupling timescale $\approx 22$ Myr for a $1\, M_\odot$ star. 

Our results are a promising step towards more physically motivated gyrochronology relations, and highlight the necessity to take into account the internal transport of angular momentum in modeling the rotational evolution of solar-like stars.
In addition, our model provides constraints on the currently unknown processes that transport angular momentum in the interior of solar-like stars.

\begin{acknowledgements}
FS is supported by the German space agency (Deutsches Zentrum f\"ur Luft- und Raumfahrt) under PLATO Data Center grant 50OO1501.
\end{acknowledgements}

\bibliographystyle{aa}

\begin{thebibliography}{}

\bibitem[Barnes(2003)]{Barnes:2003} Barnes, S.~A.\ 2003, \apj, 586, 464

\bibitem[Barnes(2007)]{Barnes:2007} Barnes, S.~A.\ 2007, \apj, 669, 1167

\bibitem[Barnes(2010)]{Barnes:2010} Barnes, S.~A.\ 2010, \apj, 722, 222

\bibitem[Barnes \& Kim(2010)]{Barnes_Kim:2010} Barnes, S.~A. \& Kim, Y.-C.\ 2010, \apj, 721, 675

\bibitem[Barnes et al.(2016a)]{Barnes_ea:2016a} Barnes, S.~A., Weingrill, J., Fritzewski, D., et al.\ 2016a, \apj, 823, 16

\bibitem[Barnes et al.(2016b)]{Barnes_ea:2016b} Barnes, S.~A., Spada, F., \& Weingrill, J.\ 2016b, Astronomische Nachrichten, 337, 810

\bibitem[Bouvier et al.(2014)]{Bouvier_ea:2014} Bouvier, J., Matt, S.~P., Mohanty, S., et al.\ 2014, Protostars and Planets VI, 433

\bibitem[Brown(2014)]{Brown:2014} Brown, T.~M.\ 2014, \apj, 789, 101

\bibitem[Brun et al.(2011)]{Brun_ea:2011} Brun, A.~S., Miesch, M.~S., \& Toomre, J.\ 2011, \apj, 742, 79

\bibitem[Charbonneau \& MacGregor(1993)]{Charbonneau_MacGregor:1993} Charbonneau, P. \& MacGregor, K.~B.\ 1993, \apj, 417, 762

\bibitem[Curtis et al.(2019)]{Curtis_ea:2019} Curtis, J.~L., Ag{\"u}eros, M.~A., Douglas, S.~T., et al.\ 2019, \apj, 879, 49

\bibitem[Demarque \& Larson(1964)]{Demarque_Larson:1964} Demarque, P.~R. \& Larson, R.~B.\ 1964, \apj, 140, 544

\bibitem[Demarque et al.(2008)]{Demarque_ea:2008} Demarque, P., Guenther, D.~B., Li, L.~H., et al.\ 2008, \apss, 316, 31

\bibitem[Denissenkov et al.(2010)]{Denissenkov_ea:2010} Denissenkov, P.~A., Pinsonneault, M., Terndrup, D.~M., et al.\ 2010, \apj, 716, 1269

\bibitem[Douglas et al.(2019)]{Douglas_ea:2019} Douglas, S.~T., Curtis, J.~L., Ag{\"u}eros, M.~A., et al.\ 2019, \apj, 879, 100

\bibitem[Eggenberger et al.(2019)]{Eggenberger_ea:2019} Eggenberger, P., Buldgen, G., \& Salmon, S.~J.~A.~J.\ 2019, \aap, 626, L1

\bibitem[Endal \& Sofia(1981)]{Endal_Sofia:1981} Endal, A.~S., \& Sofia, S.\ 1981, \apj, 243, 625

\bibitem[Gallet \& Bouvier(2015)]{Gallet_Bouvier:2015} Gallet, F., \& Bouvier, J.\ 2015, \aap, 577, A98

\bibitem[Gondoin(2017)]{Gondoin:2017} Gondoin, P.\ 2017, \aap, 599, A122

\bibitem[Hern{\'a}ndez et al.(2008)]{Hernandez_ea:2008} Hern{\'a}ndez, J., Hartmann, L., Calvet, N., et al.\ 2008, \apj, 686, 1195

\bibitem[Kawaler(1988)]{Kawaler:1988} Kawaler, S.~D.\ 1988, \apj, 333, 236

\bibitem[Koenigl(1991)]{Koenigl:1991} Koenigl, A.\ 1991, \apjl, 370, L39

\bibitem[Kippenhahn et al.(2012)]{KWW12}  Kippenhahn R., Weigert A., Weiss A. \ 2012, Stellar Structure and Evolution, 2nd edn. Springer-Verlag Berlin Heidelberg

\bibitem[Kraft(1967)]{Kraft:1967} Kraft, R.~P.\ 1967, \apj, 150, 551

\bibitem[Lanzafame \& Spada(2015)]{LS15} Lanzafame, A.~C. \& Spada, F.\ 2015, \aap, 584, A30

\bibitem[MacGregor \& Brenner(1991)]{MacGregor_Brenner:1991} MacGregor, K.~B. \& Brenner, M.\ 1991, \apj, 376, 204

\bibitem[Meibom et al.(2009)]{Meibom_ea:2009} Meibom, S., Mathieu, R.~D., \& Stassun, K.~G.\ 2009, \apj, 695, 679

\bibitem[Meibom et al.(2011)]{Meibom_ea:2011} Meibom, S., Mathieu, R.~D., Stassun, K.~G., et al.\ 2011, \apj, 733, 115

\bibitem[Moraux et al.(2013)]{Moraux_ea:2013} Moraux, E., Artemenko, S., Bouvier, J., et al.\ 2013, \aap, 560, A13

\bibitem[Nelder \& Mead(1965)]{Nelder_Mead:1965} Nelder, J. A., \& Mead, R. \ 1965, Comput. J., 7, 308

\bibitem[Newton et al.(2018)]{Newton_ea:2018} Newton, E.~R., Mondrik, N., Irwin, J., et al.\ 2018, \aj, 156, 217

\bibitem[Oglethorpe \& Garaud(2013)]{Oglethorpe_Garaud:2013} Oglethorpe, R.~L.~F., \& Garaud, P.\ 2013, \apj, 778, 166

\bibitem[Pinsonneault et al.(1989)]{Pinsonneault_ea:1989} Pinsonneault, M.~H., Kawaler, S.~D., Sofia, S., et al.\ 1989, \apj, 338, 424

\bibitem[Rebull(2001)]{Rebull:2001} Rebull, L.~M.\ 2001, \aj, 121, 1676

\bibitem[Rebull et al.(2016)]{Rebull_ea:2016} Rebull, L.~M., Stauffer, J.~R., Bouvier, J., et al.\ 2016, \aj, 152, 113

\bibitem[Ruediger \& Kitchatinov(1996)]{Ruediger_Kitchatinov:1996} Ruediger, G. \& Kitchatinov, L.~L.\ 1996, \apj, 466, 1078

\bibitem[Schatzman(1962)]{Schatzman:1962} Schatzman, E.\ 1962, Annales d'Astrophysique, 25, 18

\bibitem[Skumanich(1972)]{Skumanich:1972} Skumanich, A.\ 1972, \apj, 171, 565

\bibitem[Soderblom et al.(1993)]{Soderblom_ea:1993} Soderblom, D.~R., Stauffer, J.~R., Hudon, J.~D., et al.\ 1993, \apjs, 85, 315

\bibitem[Somers \& Pinsonneault(2016)]{Somers_Pinsonneault:2016} Somers, G. \& Pinsonneault, M.~H.\ 2016, \apj, 829, 32

\bibitem[Spada et al.(2010)]{Spada_ea:2010} Spada, F., Lanzafame, A.~C., \& Lanza, A.~F.\ 2010, \mnras, 404, 641

\bibitem[Spada et al.(2017)]{Spada_ea:2017} Spada, F., Demarque, P., Kim, Y.-C., et al.\ 2017, \apj, 838, 161

\bibitem[Spruit(2002)]{Spruit:2002} Spruit, H.~C.\ 2002, \aap, 381, 923

\bibitem[Stauffer et al.(1984)]{Stauffer_ea:1984} Stauffer, J.~R., Hartmann, L., Soderblom, D.~R., et al.\ 1984, \apj, 280, 202

\bibitem[Stauffer et al.(1985)]{Stauffer_ea:1985} Stauffer, J.~R., Hartmann, L.~W., Burnham, J.~N., et al.\ 1985, \apj, 289, 247

\bibitem[Talon \& Charbonnel(2003)]{Talon_Charbonnel:2003} Talon, S., \& Charbonnel, C.\ 2003, \aap, 405, 1025

\bibitem[Charbonnel \& Talon(2005)]{Talon_Charbonnel:2005} Charbonnel, C., \& Talon, S.\ 2005, Science, 309, 2189

\bibitem[Weber \& Davis(1967)]{Weber_Davis:1967} Weber, E.~J. \& Davis, L.\ 1967, \apj, 148, 217

\end{thebibliography}

\newpage

\appendix

\section{Rotational isochrones}
\label{appendix}

\begin{figure}
\begin{center}
\includegraphics[width=0.5\textwidth]{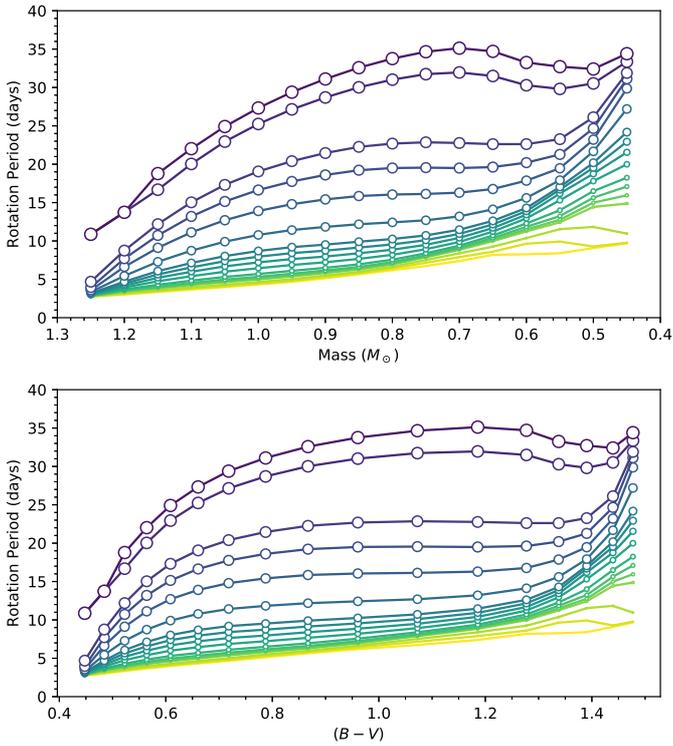}
\caption{Rotational isochrones reported in Table \ref{tab:iso} (age range: $100$ Myr to $4.57$ Gyr). Top panel: rotation period vs. mass; bottom: rotation period vs. $(B-V)$ color. The size of the circles increases monotonically with age.}
\label{fig:isoplot}
\end{center}
\end{figure}

Selected rotational isochrones calculated with our updated two-zone model are reported in Table \ref{tab:iso}, and plotted in Figure \ref{fig:isoplot}.
The Table lists the surface rotation period, in days, as a function of stellar mass and age; $(B-V)$ colors are also given, as calculated from a solar metallicity, $400$ Myr isochrone from the YaPSI database \citep{Spada_ea:2017}.

\begin{figure}
\begin{center}
\includegraphics[width=0.5\textwidth]{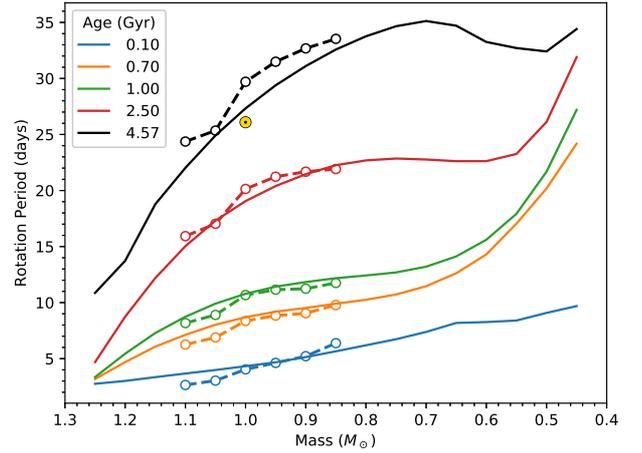}
\caption{Rotational isochrones from this work (solid lines) compared with those of \citetalias{LS15} (dashed lines with circles). The Sun is also shown for reference.}
\label{fig:ls15iso}
\end{center}
\end{figure}

In Figure \ref{fig:ls15iso} we compare our current isochrones with those of \citetalias{LS15}.
The updated isochrones are consistent with our previous results in the range of overlap.
A moderate disagreement is visible at  ages $\gtrsim 2$ Gyr, reflecting the different choice of the overall calibration constant $K_w$ in equation \eqref{wind_braking}.

\begin{sidewaystable*}
\caption{Rotational isochrones constructed with our updated two-zone model; surface rotation period is reported in days.}
\begin{center}
\begin{tabular}{cc|cccccccccccccccccccc}
\hline
\hline
Mass & $(B-V)$ & \multicolumn{17}{c}{Ages (Gyr)} \\
&  &     0.10 &     0.12 &     0.15 &     0.20 &     0.22 &     0.25 &     0.30 &     0.40 &     0.50 &     0.60 &     0.70 &     1.00 &     1.50 &     2.00 &     2.50 &     4.00 &     4.57 \\ 
\hline 
    0.45 &    1.477 &   9.69 &   9.76 &  10.96 &  14.87 &  15.95 &  17.09 &  18.24 &  19.99 &  21.54 &  22.92 &  24.18 &  27.20 &  29.87 &  31.13 &  31.89 &  33.36 &  34.40 \\ 
    0.50 &    1.440 &   9.08 &   9.28 &  11.83 &  14.43 &  14.99 &  15.63 &  16.50 &  17.80 &  18.76 &  19.53 &  20.19 &  21.68 &  23.20 &  24.66 &  26.12 &  30.53 &  32.41 \\ 
    0.55 &    1.390 &   8.40 &   9.92 &  11.54 &  12.43 &  12.76 &  13.24 &  14.00 &  15.30 &  16.18 &  16.69 &  17.04 &  17.92 &  19.47 &  21.27 &  23.26 &  29.81 &  32.71 \\ 
    0.60 &    1.338 &   8.26 &   9.65 &  10.38 &  11.25 &  11.50 &  11.84 &  12.28 &  12.94 &  13.43 &  13.88 &  14.30 &  15.61 &  17.86 &  20.20 &  22.62 &  30.29 &  33.25 \\ 
    0.65 &    1.277 &   8.18 &   8.60 &   9.25 &   9.97 &  10.16 &  10.39 &  10.71 &  11.22 &  11.68 &  12.15 &  12.62 &  14.12 &  16.77 &  19.62 &  22.62 &  31.50 &  34.71 \\ 
    0.70 &    1.186 &   7.38 &   7.89 &   8.38 &   8.84 &   8.96 &   9.12 &   9.38 &   9.87 &  10.40 &  10.92 &  11.46 &  13.20 &  16.29 &  19.50 &  22.77 &  31.96 &  35.12 \\ 
    0.75 &    1.072 &   6.73 &   7.15 &   7.51 &   7.87 &   7.97 &   8.12 &   8.39 &   8.92 &   9.49 &  10.10 &  10.72 &  12.69 &  16.13 &  19.55 &  22.86 &  31.74 &  34.66 \\ 
    0.80 &    0.961 &   6.19 &   6.43 &   6.64 &   6.94 &   7.06 &   7.24 &   7.56 &   8.19 &   8.86 &   9.55 &  10.25 &  12.42 &  16.05 &  19.50 &  22.69 &  31.02 &  33.76 \\ 
    0.85 &    0.867 &   5.65 &   5.79 &   5.98 &   6.29 &   6.41 &   6.61 &   6.95 &   7.66 &   8.39 &   9.14 &   9.90 &  12.18 &  15.85 &  19.22 &  22.26 &  30.02 &  32.58 \\ 
    0.90 &    0.787 &   5.14 &   5.31 &   5.50 &   5.82 &   5.95 &   6.15 &   6.50 &   7.22 &   7.97 &   8.73 &   9.52 &  11.83 &  15.41 &  18.60 &  21.45 &  28.70 &  31.10 \\ 
    0.95 &    0.718 &   4.67 &   4.81 &   5.01 &   5.36 &   5.51 &   5.73 &   6.11 &   6.87 &   7.64 &   8.41 &   9.19 &  11.42 &  14.79 &  17.75 &  20.40 &  27.15 &  29.39 \\ 
    1.00 &    0.661 &   4.33 &   4.45 &   4.64 &   4.98 &   5.12 &   5.33 &   5.70 &   6.45 &   7.21 &   7.95 &   8.69 &  10.79 &  13.91 &  16.64 &  19.06 &  25.24 &  27.33 \\ 
    1.05 &    0.609 &   3.99 &   4.10 &   4.29 &   4.62 &   4.75 &   4.96 &   5.31 &   5.99 &   6.69 &   7.36 &   8.02 &   9.91 &  12.70 &  15.14 &  17.31 &  22.94 &  24.91 \\ 
    1.10 &    0.564 &   3.67 &   3.78 &   3.96 &   4.25 &   4.36 &   4.53 &   4.82 &   5.40 &   5.99 &   6.56 &   7.13 &   8.74 &  11.12 &  13.19 &  15.03 &  20.03 &  22.02 \\ 
    1.15 &    0.523 &   3.35 &   3.45 &   3.58 &   3.81 &   3.90 &   4.05 &   4.28 &   4.74 &   5.20 &   5.63 &   6.07 &   7.29 &   9.11 &  10.72 &  12.19 &  16.68 &  18.78 \\ 
    1.20 &    0.484 &   3.01 &   3.09 &   3.21 &   3.40 &   3.46 &   3.55 &   3.69 &   3.94 &   4.18 &   4.44 &   4.69 &   5.44 &   6.62 &   7.69 &   8.74 &  13.73 &  *** \\ 
    1.25 &    0.448 &   2.76 &   2.77 &   2.78 &   2.82 &   2.83 &   2.86 &   2.89 &   2.97 &   3.04 &   3.11 &   3.18 &   3.35 &   3.58 &   3.94 &   4.69 &  10.87 &  *** \\ 
\hline
\end{tabular}
\tablefoot{In the isochrone at $4.57$ Gyr, missing entries (indicated with `***') correspond to stars that have already left the main sequence by that age.}
\end{center}
\label{tab:iso}
\end{sidewaystable*}

\end{document}